\documentclass{ws-procs9x6}

\begin{document}

\title{Masses and magnetic moments of ground-state  baryons in covariant baryon chiral perturbation theory}

\author{L. S. Geng$^*$}

\address{Research Center for Nuclear Science and Technology and
School of Physics and Nuclear Energy Engineering, Beihang University, Beijing 100191, China}

\author{J. Martin Camalich}
\address{Department of Physics and Astronomy, University of Sussex, BN1 9QH, Brighton, UK}

\author{L. Alvarez-Ruso and M. J. Vicente-Vacas}

\address{
 Departamento de F\'{\i}sica Te\'orica and IFIC, Universidad de
Valencia-CSIC, E-46071 Valencia, Spain}

\begin{abstract}
We report on some recent developments in our understanding of the light-quark mass dependence
and the SU(3) flavor symmetry breaking corrections to the magnetic moments of the ground-state baryons in
a covariant formulation of baryon chiral perturbation theory, the so-called EOMS formulation. We show that
this covariant ChPT exhibits some promising features compared to its heavy-baryon and infrared counterparts.
\end{abstract}

\keywords{mass; magnetic moment; ground-state baryons; chiral perturbation theory}

\bodymatter

\section{Introduction}\label{aba:sec1}

Effective field theories provide a way to understand low-energy strong-interaction phenomena in a systematic and controlled manner. In particular,
chiral perturbation theory (ChPT) has long been deemed as the low-energy effective field theory of QCD. It has been
widely used in studying many properties of hadrons (see Ref.~\cite{Scherer:2012zzd} and references therein). Historically, applications of
baryon ChPT have suffered from the so-called power-counting-breaking (PCB) problem, which is due to the large non-zero baryon masses in the
chiral limit~\cite{Gasser:1987rb}, and different formulations to deal with
 this problem have been proposed, including the
heavy-baryon ChPT~\cite{Jenkins:1990jv}, infrared ChPT~\cite{Becher:1999he} and its modified version~\cite{Gail-2007-PhdThesis}, and the extended-on-mass-shell (EOMS) ChPT~\cite{Fuchs:2003qc}.

Over the past few years, the EOMS formulation of baryon ChPT has been successfully applied
to study a number of physical observables~\cite{Geng:2008mf,Geng:2009hh,Geng:2009ik,Geng:2009ys,MartinCamalich:2010fp,Geng:2011wq,Alarcon:2011zs,Alarcon:2011px,MartinCamalich:2011py} (see also Ref.~\cite{Pascalutsa:2011fp} and references cited therein). Compared to the more conventional HB and infrared formulations, the EOMS formulation not only satisfies all the symmetry constraints but also
converges relatively faster, particularly in the three flavor sector and at large unphysical light quark masses in the two flavor space.
In this talk we briefly introduce two recent works studying the magnetic moments and
the masses of the lowest-lying octet/decuplet baryons.~\footnote{These quantities have also been
studied in the finite-range regularized ChPT. For details, see Refs.~\cite{Young:2009zb,Hall:2012pk}.}

\section{Magnetic moments}

The magnetic moments (MMs) of the octet baryons have long been
related to those of the proton and neutron, i.e., the celebrated
Coleman-Glashow (CG) relations~\cite{Coleman:1961jn}. These
relations are a result of (approximate) global SU(3) flavor
symmetry. It was soon realized that ChPT may be employed to study SU(3)
breaking effects on the MMs of the baryon octet.  The first effort
was undertaken by Caldi and Pagels in 1974~\cite{Caldi:1974ta}, even
before modern ChPT was formulated. It was found that at
next-to-leading-order (NLO), SU(3) breaking effects are so large
that the description of the octet baryon MMs by the CG relations
tends to deteriorate, which was later confirmed by the calculations
performed in Heavy Baryon (HB)
ChPT~\cite{Jenkins:1992pi,Durand:1997ya,Puglia:1999th,Meissner:1997hn}
and Infrared (IR) ChPT~\cite{Kubis:2000aa}. This apparent failure
has often been used to question the validity of SU(3) ChPT in the
one-baryon sector.  In order to solve this problem, different
approaches have been suggested, including reordering the chiral
series~\cite{Mojzis:1999qw}
 or using a cutoff to reduce the loop contributions, i.e., the so-called long-range regularization~\cite{Donoghue:1998bs}.

We have shown that the above-mentioned apparent failure
of baryon SU(3) ChPT is caused by the power-counting-restoration (PCR) procedure used in removing the PCB terms.
In the following, we discuss the results for the octet~\footnote{For the
study of the decuplet baryons in the EOMS formulation of baryon ChPT see Ref.~\cite{Geng:2009ys}.} baryon
magnetic moments at NLO without considering the contributions of
dynamical decuplet baryons, whose effects can be found in Ref.~\cite{Geng:2009hh}.

\begin{figure}

\begin{center}
\includegraphics[width=7cm]{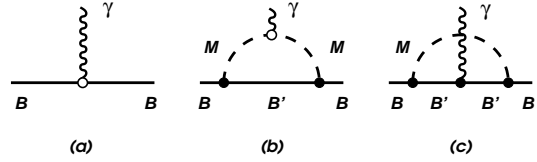}
\caption{\label{fig1}   Feynman diagrams contributing to the
octet baryon magnetic moments up to NLO.}
\end{center}
\end{figure}

Up to NLO, one has the diagrams shown in Fig.~\ref{fig1}.
 The tree-level coupling
\textbf{\textit{(a)}} gives the leading-order (LO) result
\begin{equation}
\kappa_B^{(2)}=\alpha_B b_6^D+\beta_B b_6^F, \label{eq:treeL}
\end{equation}
where the coefficients $\alpha_B$ and $\beta_B$ for each of the octet baryons
 are listed in Table I of Ref.~\cite{Geng:2008mf}. This lowest-order contribution is
nothing  but the SU(3)-symmetric prediction leading to the CG
relations ~\cite{Coleman:1961jn, Jenkins:1992pi}.

The $\mathcal{O}(p^3)$ diagrams \textbf{\textit{(b)}} and
\textbf{\textit{(c)}} account for the leading SU(3)-breaking
corrections that are induced by the corresponding degeneracy
breaking in the masses of the pseudoscalar meson octet. Their
contributions to the anomalous magnetic moment of a given member of
the octet $B$ can be written as
\begin{eqnarray}
\kappa^{(3)}_B&=&\frac{1}{8\pi^2 F_\phi^2}\left(\sum_{M=\pi,K}\xi_{BM}^{(b)}
H^{(b)}(m_M)+\sum_{M=\pi,K,\eta}\xi_{BM}^{(c)} H^{(c)}(m_M)\right)\label{eq:thirdO}
\end{eqnarray}
with the coefficients $\xi_{BM}^{(b,c)}$ listed in Table I of Ref.~\cite{Geng:2008mf}. The
loop-functions, which are convergent, read as
\begin{eqnarray}
 H^{(b)}(m)&=&-M_B^2+2 m^2+\frac{m^2}{M_B^2}(2
M_B^2-m^2)\log\left(\frac{m^2}{M_B^2}\right)\nonumber \\
&+&\frac{2m\left(m^4-4 m^2 M_B^2+2
M_B^4\right)}{M_B^2\sqrt{4M_B^2-m^2}}\,\arccos\left(\frac{m}{2\,M_B}\right),
\nonumber \\
H^{(c)}(m)&=&M_B^2+2m^2+\frac{m^2}{M_B^2}(M_B^2-m^2)\log\left(\frac{m^2}{M_B^2}
\right)\nonumber \\
&+&\frac{2m^3\left(m^2-3
M_B^2\right)}{M_B^2\sqrt{4M_B^2-m^2}}\,\arccos\left(\frac{m}{2\,M_B}\right).
\label{eq:loop}
\end{eqnarray}
 One immediately notices that they contain pieces $\sim
M_B^2$ that contribute at $\mathcal{O}(p^2)$ to the MMs, which break
the naive PC. These terms have to be removed by applying a PCR scheme, such as the HB, the infrared, or the EOMS schemes.

In Table \ref{table2}, we show the LO and NLO results obtained in
the EOMS scheme~\cite{Geng:2008mf}. For the sake of comparison, we
also show the NLO results obtained by using the HB and IR schemes.
To compare with the results of earlier studies, we define
\begin{equation}
 \tilde{\chi}^2=\sum (\mu_\mathrm{th}-\mu_\mathrm{exp})^2,
\end{equation}
 while $\mu_\mathrm{th}$ and $\mu_\mathrm{exp}$ are theoretical
and experimental MMs of the octet baryons. The results shown in
Table \ref{table2} are obtained by minimizing $\tilde{\chi}^2$ with
respect to the two LECs $\tilde{b}^D_6$ and $\tilde{b}^F_6$,
renormalized $b^D_6$ and $b^F_6$. It is clear that the HB and IR
results spoil the CG relations, as found in previous studies, while
the EOMS results improve them.

\begin{table}[t]
\scriptsize
\tablefont
\tbl{The baryon-octet magnetic moments (in nuclear magnetons)
up to $\mathcal{O}(p^3)$ obtained in different $\chi$PT approaches
in comparison with data. \label{table2}}
{\begin{tabular*}{160mm}{ccccccccccc}
\toprule
 & $p$ & $n$ & $\Lambda$ & $\Sigma^-$ & $\Sigma^+$ & $\Sigma^0$ & $\Xi^-$ &
$\Xi^0$ & $\Lambda\Sigma^0$ &
$\tilde{\chi}^2$ \\
\hline
\multicolumn{11}{c}{$\mathcal{O}(p^2)$}  \\
\hline
Tree level & 2.56 & -1.60 & -0.80 & -0.97 & 2.56 & 0.80 & -1.60 & -0.97 & 1.38 & 0.46  \\
\hline
\multicolumn{11}{c}{$\mathcal{O}(p^3)$}  \\
\hline
HB  & 3.01 & -2.62 & -0.42 & -1.35 & 2.18 & 0.42 & -0.70 & -0.52 & 1.68 & 1.01  \\

IR & 2.08 & -2.74 & -0.64 & -1.13 & 2.41 & 0.64 & -1.17 & -1.45 & 1.89 & 1.86 \\

EOMS & 2.58 & -2.10 & -0.66 & -1.10 & 2.43 & 0.66 & -0.95 & -1.27 & 1.58 & 0.18  \\
\hline Exp. &  2.793(0) & -1.913(0) & -0.613(4) & -1.160(25) &
2.458(10) & --- &
-0.651(3) &-1.250(14) & $\pm$  1.61(8) &   \\
\botrule
\end{tabular*}}
\end{table}

\begin{figure}
\begin{center}
\includegraphics[width=8cm]{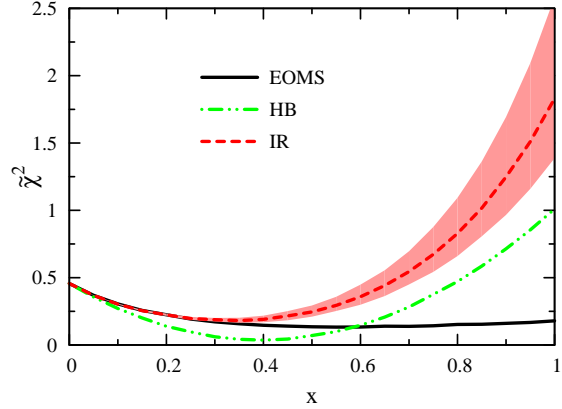}
\caption{\label{fig2}   SU(3)-breaking evolution of the minimal
$\tilde{\chi}^2$ in the $\mathcal{O}(p^3)$ $\chi$PT approaches under
study. The shaded bands are produced by
 varying $M_B$ from 0.8 to 1.1 GeV.}
\end{center}
\end{figure}

The difference between the EOMS, HB, and IR approaches can also be
seen from Fig.~\ref{fig2}, where we show the evolution of the
minimal $\tilde{\chi}^2$ as a function of $x=M_M/M_{M,phys}$, while
$M_M$, $M_{M,phys}$ are the masses of the pion, kaon, eta used in
the calculation and their physical values. It is clear that at
$x=0$, the chiral limit, all the results are identical to the CG
relations. As $x$ approaches 1, where the meson masses equal to the
physical values, only the EOMS results show a proper behavior, while
both the HB and IR results rise sharply. This shows clearly that
relativity and analyticity of the loop results play an important
role in the present case.
\section{Octet Baryon Masses}
In the last few years, a number of lattice QCD collaborations reported
the results of 2+1 and 2 flavor simulations of the octet and decuplet baryon masses, including the BMW~\cite{Durr:2008zz},
the PACS-CS~\cite{Aoki:2008sm},
the HSC~\cite{Lin:2008pr}, the LHP~\cite{WalkerLoud:2008bp}, and the ETM~\cite{Alexandrou:2009qu} collaborations.
Surprisingly, when trying to describe their simulations with the NLO HB ChPT, both
the PACS and LHP collaborations realized that they cannot obtain reasonable fits unless they
allow some of the LECs, e.g.,
$C$, $D$, and $F$ (with their definitions given below), to almost vanish, inconsistent with their empirical values, ~\cite{WalkerLoud:2008bp,Ishikawa:2009vc}. The findings
are rather unexpected given the fact that in the two flavor sector, ChPT seems to work rather well, at least the IR formulation (see, e.g.,
Refs.~\cite{Procura:2003ig,Procura:2006bj}). To understand this situation, we performed a first
NLO study of the octet and decuplet baryon masses employing the EOMS scheme~\cite{MartinCamalich:2010fp}.

\begin{figure}[htpb]
\begin{center}
\includegraphics[width=8cm]{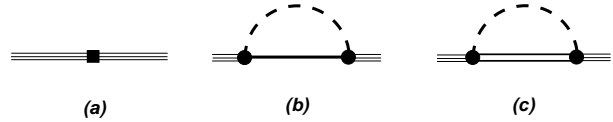}
\caption{Feynman diagrams contributing to the octet- and decuplet-baryon masses ($B$ and $T$ respectively) up to
$\mathcal{O}(p^3)$ in $\chi$PT. The solid lines correspond to octet-baryons, double lines to decuplet-baryons and dashed
lines to mesons. The black dots indicate $1^{st}$-order couplings while boxes, $2^{nd}$-order couplings.
\label{Fig:masses}}
\end{center}
\end{figure}
At $\mathcal{O}(p^2)$ the following terms in the chiral Lagrangian contribute to the octet and decuplet masses
\begin{eqnarray}
&&\mathcal{L}_{B}^{(2)}=b_0 \langle \chi_+\rangle\langle\bar{B}B\rangle+b_{D/F}\langle \bar{B}[\chi_{+},B]_\pm\rangle, \label{Eq:OctetCT}\\
&&\mathcal{L}_{T}^{(2)}=\frac{t_0}{2}\bar{T}^{abc}_\mu g^{\mu\nu}T_\nu^{abc}\langle\chi_{+}\rangle + \frac{t_D}{2}\bar{T}^{abc}_\mu g^{\mu\nu}\left(\chi_+,T_\nu\right)^{abc}, \nonumber
\end{eqnarray}
where $\langle X\rangle$ is the trace in flavor space and $(X,T_\mu)^{abc}\equiv(X)_d^a T_\mu^{dbc}+(X)_d^b T_\mu^{adc}+(X)_d^c T_\mu^{abd}$.
In the Eqs. (\ref{Eq:OctetCT}), $\chi_+$ introduces the explicit chiral symmetry breaking, and
the coefficients $b_0$, $b_D$, $b_F$, and $t_0$, $t_D$ are unknown LECs.

For the calculation of the leading loop contributions of Fig.~\ref{Fig:masses} to the masses with pseudoscalar
mesons ($\phi$), octet- ($B$) and decuplet-baryons ($T$) we use
the lowest-order $\phi B$ Lagrangian and the $\phi BT$ and $\phi T$ ones of Refs.~\cite{Geng:2009hh,Geng:2009ys}.
We take the empirical values $D=0.80$ and $F=0.46$ for the $\phi B$ couplings, and
$\mathcal{C}=1.0$~\cite{Geng:2009hh} and  $\mathcal{H}=1.13$~\cite{Geng:2009ys} for the  $\phi B T$ and $\phi T$ ones, instead of
treating them as free parameters as in the LQCD analyses~\cite{WalkerLoud:2008bp,Ishikawa:2009vc}.

We calculate the loops in the covariant approach and recover the power-counting using the EOMS renormalization
prescription~\cite{Fuchs:2003qc}. From the covariant
results we have obtained the HB ones by defining $M_D=M_B+\delta$ and expanding the loop-functions about the limit
$M_B\rightarrow \infty$.

 In the following, we extend our analysis to the 2+1-flavor lQCD results of the
PACS-CS collaboration~\cite{Aoki:2008sm}. We choose the points for which both the pion and kaon masses are below
600 MeV and perform fits of the 7 parameters, $M_{B0}$, $b_0$, $b_D$, $b_F$, $M_{D0}$, $t_0$ and $t_D$, to the chosen
24 lattice points that we assume to have independent statistical errors ($\sigma_i$) but fully-correlated errors propagated
from the determination of the lattice spacing. The
finite volume corrections have been included~\cite{Geng:2011wq}.

The results of the fits  are shown in Table~\ref{Table:ResChExtrapolation}. The error bar quoted in the GMO and HB columns and the first one assigned to the covariant results are the uncertainties propagated from the fitted parameters. The second error bar in the covariant results is a theoretical uncertainty coming from the truncation of the chiral expansion which is estimated by taking 1/2 of the difference between the results obtained at LO and NLO.

Clearly the NLO EOMS ChPT describes much better the ligh-quark mass dependence of the octet baryon masses than the NLO
HB ChPT. On the other hand, one need to study whether at NNLO the EOMS ChPT works still better than its HB (or IR) counterpart.
Such a work is in progress. It should be noted that recently Semke and Lutz have made such an attempt using their formulation. With the help of
a large-$N_c$ operator analysis to relate the large number of LECs, they were able to describe reasonably well the latest lattice QCD simulations~\cite{Semke:2011ez}.

\begin{table}
\centering
\tbl{Extrapolation in MeV  from the fits to the PACS-CS results~\cite{Aoki:2008sm} on the baryon masses using B$\chi$PT up to
NLO. The $\chi^2$ is the estimator for the fits to the lQCD results whereas $\bar{\chi}^2$ include also experimental data. See Ref.~\cite{MartinCamalich:2010fp} for details. \label{Table:ResChExtrapolation}}
{\tablefont
\begin{tabular}{ccccc}\toprule
&GMO&HB&Covariant&Expt.\\
\hline
$M_N$&$971(22)$&$764(21)$&$893(19)(39)$&$940(2)$\\
$M_\Lambda$&$1115(21)$&$1042(20)$&$1088(20)(14)$&$1116(1)$\\
$M_\Sigma$&$1165(23)$&$1210(22)$&$1178(24)(7)$&$1193(5)$\\
$M_\Xi$&$1283(22)$&$1392(21)$&$1322(24)(20)$&$1318(4)$\\

$M_\Delta$&$1319(28)$&$1264(22)$&$1222(24)(49)$&$1232(2)$\\
$M_{\Sigma^*}$&$1433(27)$&$1466(22)$&$1376(24)(29)$&$1385(4)$\\
$M_{\Xi^*}$&$1547(27)$&$1622(23)$&$1531(25)(8)$&$1533(4)$\\
$M_{\Omega^-}$&$1661(27)$&$1733(25)$&$1686(28)(13)$&$1672(1)$\\
\hline
$\chi^2_{\rm d.o.f.}$& $0.63$ &$9.2$&$2.1$ \\
$\bar{\chi}^2_{\rm d.o.f.}$& $4.2$ &$36.6$&$2.8$ \\\botrule
\end{tabular}}
\end{table}
\section{Summary and outlook}
In the past few years, the EOMS formulation of baryon ChPT have been successfully applied
to study a number of important physical quantities. Recently, the EOMS scheme
has also been utilized to formulate a covariant ChPT for heavy-light systems and the results
are very encouraging~\cite{Geng:2010vw,Geng:2010df,Altenbuchinger:2011qn}. We expect to see more applications of covariant baryon ChPT or heavy-meson ChPT
in the near future.

\section{Acknowledgement}
L.S. Geng acknowledges support from the Fundamental Research Funds for the Central Universities and
 the National Natural
Science Foundation of China (Grant No. 11005007). M. J. Vicente-Vacas
was supported by DGI and FEDER funds, under contracts FIS2011-28853-C02-01,
by Generalitat Valenciana contract PROMETEO/2009/0090 and by the EU Hadron-
Physics2 project, grant agreement no. 227431.

\bibliographystyle{ws-procs9x6}
\bibliography{ws-pro-sample}

\end{document}